\documentclass[12pt]{iopart}

\usepackage{iopams}
\usepackage{graphicx}
\usepackage{harvard}

\usepackage[active]{srcltx}  %SRC Specials for inverse DVI search

\begin{document}

\title[Short-time homomorphic wavelet estimation]{Short-time homomorphic wavelet estimation}

\author{Roberto Henry Herrera and Mirko van der Baan}

\address{Department of Physics, University of Alberta,
 Edmonton T6G 2E1, CA}
\eads{rhherrer@ualberta.ca, Mirko.VanderBaan@ualberta.ca}
\begin{abstract}
Successful wavelet estimation is an essential step for seismic methods like impedance inversion, analysis of amplitude variations with offset and full waveform inversion. Homomorphic deconvolution has long intrigued as a potentially elegant solution to the wavelet estimation problem. Yet a successful implementation has proven difficult. Associated disadvantages like phase unwrapping and restrictions of sparsity in the reflectivity function limit its application. We explore short-time homomorphic wavelet estimation as a combination of the classical homomorphic analysis and log-spectral averaging. The introduced method of log-spectral averaging using a short-term Fourier transform increases the number of sample points, thus reducing estimation variances. We apply the developed method on synthetic and real data examples and demonstrate good performance.

%Statistical methods are now in a sufficiently mature stage to be used as standalone tools for wavelet estimation.

\end{abstract}

\vspace{2pc}

\submitto{ \emph{J. Geophys. Eng.}}
\maketitle

\section{Introduction}
Homomorphic wavelet estimation has been popular since its first application to seismic deconvolution  \cite{Ulrych1971}. In part this is due to the conceptual simplicity of the method requiring only a forward Fourier transform, then logarithm, followed by an inverse Fourier transform. In addition, no minimum phase assumptions are made on the wavelet, nor is the reflectivity assumed white in the original formulation \cite{Ulrych1971}.

Despite its promise, the method has often produced mixed results \cite{Jin1984}. This is partially due to an often overlooked condition, namely that the reflectivity series must be sufficiently sparse \cite{Ulrych1971}. Also the method forces the resulting wavelet to replicate the strongest arrival in the data, often the first arrival in data not subjected to an automatic gain control to equalize amplitudes \cite{Tribolet1978}. In this paper, we combine elements of \possessivecite{Ulrych1971} classic application of homomorphic analysis, \possessivecite{Tribolet1978} short-time cepstral deconvolution with homomorphic wavelet estimation in the log-spectral domain \cite{Otis1977}. This relaxes some of the original assumptions; for instance, the assumption of a sufficiently sparse reflectivity is replaced by a white, random one. The minimum reflectivity assumption is replaced by short mixed phase segments by windowing. This produces better wavelet estimates while extending the ability to handle nonminimum phases in both the wavelet and the reflectivity.

We first describe the classical cepstral liftering method and the stacking procedure in the cepstral domain. Next we describe our approach. The new proposal is based on short-time homomorphic analysis by averaging in the log-spectral domain. We use constant length windows in the log-spectrum whereas \possessivecite{Tribolet1978} short-time approach was based on averaging different window sizes in the cepstral domain. Finally we test and validate the performance of the short-time homomorphic method on synthetic and real datasets.

\section{Method}
\subsection{The classical homomorphic analysis}
The homomorphic analysis is related to signals which are the outcomes of convolution, thus \cite{Ulrych1971}:
\begin{equation}\label{eq:convmodelorg}
s_i(t)= w(t)\star r_i(t),
\end{equation}
\noindent where $r_i(t)$ is the reflectivity of the $i_{th}$ trace in a seismic profile, $w(t)$ is the stationary wavelet, $\star$ stands for the convolution operator and $s_i(t)$ is the resulting $i_{th}$ seismic trace.

Following the classical homomorphic analysis \cite{Oppenheim1968,Opp2010}, we apply sequentially a Fourier transform ($FT$), complex natural logarithm ($\ln$) and inverse Fourier transform ($FT^{-1}$) to get a new signal in the cepstral domain. In the cepstral domain convolution is mapped into an addition:
\begin{equation}\label{eq:complexcepsorg}
\hat{s}_i(t)= \hat{w}(t)+ \hat{r}_i(t),
\end{equation}

\noindent where $\hat{s}_i(t)= FT^{-1}\{\ln[FT\{s_i(t)\}]\}$ is the complex cepstrum of $s_i(t)$ and $\hat{w}(t)$ and $\hat{r}_i(t)$ are the complex cepstra of the wavelet and the reflectivity respectively.

The homomorphism comes from transforming convolved signals into additive signals. This additive space is known as the quefrency domain, where all the terms are named by reversing the first syllable of their spectral domain analogues \cite{Otis1977}.

\citeasnoun{Ulrych1971} noticed that if the wavelet is time-invariant and has a smooth spectrum its contribution to the complex cepstrum will be located at the low quefrencies of the cepstral representation. On the other hand, a rapidly varying log-amplitude and phase spectra, associated to the reflectivity will have contributions at higher quefrency rhamonics. Thus, by short-pass liftering or high-pass liftering we can extract either the wavelet or the reflectivity function. These filters act as windowing in the quefrency domain. In addition if the reflectivity is minimum phase its complex cepstrum will be right-sided, making the separation of the additive terms easier. The third condition \citeasnoun{Ulrych1971} observed is related to the sparsity: if the interval time is larger than the wavelet cepstrum, two reflections overlapping in the time domain can be separated in the cepstral domain. In this latter condition the logarithm of the spectrum plays the role of whitening the wavelet spectrum to shorten its cepstral representation.

After windowing, the resulting signal is transformed to the time domain, that is:
\begin{equation}\label{eq:invcepstrum}
\tilde{s}_{i}(t) = FT^{-1}\{\exp(FT\{\hat{s}(t)\hat{f}(t)\})\},
\end{equation}

\noindent with $\hat{f}$ the windowing lifter and $\tilde{s}_{i}(t)$ is the estimated signal after the liftering process. Theoretically a low-cut time lifter will produce $\tilde{s}_{i}(t) = \tilde{w}(t)$ and a high-cut time lifter will give $\tilde{s}_{i}(t)=\tilde{r}_{i}(t)$. Judicious windowing can extract then the propagating wavelet if the reflectivity is sufficient sparse. Unfortunately this is rarely the case in recorded seismic data.

%\subsection{Cepstral stacking (Log-spectral averaging)}

To avoid cepstral windowing, \citeasnoun{Otis1977} proposed a technique based on the averaging of log-spectral components in an array of seismic traces. They considered the convolution of a constant source wavelet with a suite of non-stationary reflectivity series. In the cepstral domain the wavelet will be located at low quefrencies while the reflectivity will vary from trace to trace.  By averaging the complex cepstra of a seismic profile, the constant wavelet can be estimated since the variable reflectivity with zero mean will tend to zero:

\begin{equation}\label{eq:avecomplexceps}
\hat{s}(t)= \frac{1}{N}\sum_{i=1}^{N}\hat{s}_i(t) = w(t) + \frac{1}{N}\sum_{i=1}^{N}\hat{r}_i(t),
\end{equation}
\vspace{2mm}
\noindent where $i$ represents the $i_{th}$ trace in a seismic profile of $N$ traces.

Averaging in the cepstral domain is equivalent to averaging in the log-spectral domain, since the Fourier transform is a linear process. Thus,
\begin{equation}\label{eq:avelogspectrum2}
\hat{S}(f)= \hat{W}(f) + \frac{1}{N}\sum_{i=1}^{N} \hat{R}_i(f),
\end{equation}
\noindent where $\hat{S}(f)=\ln S(f)$ and $S(f)$ is the Fourier transform of $s(t)$, likewise for the wavelet and the reflectivity.

 Cepstral stacking will produce successful results when the wavelet is spatially stationary and the reflectivity series are white. The latter condition implies that the geological structure changes at each shot point \cite{TriboletBook}. This can be best realized by combining traces from different parts of the 3D volume. The minimum phase reflectivity assumption is maintained in the log-spectral averaging method.

\subsection{Our approach}
Spectral estimation can be improved by using the Welch approach, that is, by assuming the signal of interest is stationary, and then averaging the individual spectra obtained from partially overlapping segments taken from the total observed data \cite{TriboletBook}. The complex cepstrum of the $i_{th}$ trace is then:

\begin{equation}\label{eq4:convmodel}
\hat{s}_{ik}(t)= \hat{w}(t)+ \hat{r}_{ik}(t),    \ \ 0 \leq t \leq L,
\end{equation}

\noindent where $k$ represents the segment of length $L$ used to compute the complex cepstrum.

If the cepstral structure $\hat{r}_{ik}(t)$ is independent between segments, the estimated wavelet in the log-spectrum is:

\begin{equation}\label{eq6:logspectral}
\hat{W}_{e}(f)=\frac{1}{NM}\sum_{i=1}^{N}\sum_{k=1}^{M}\hat{S}_{ik}(f)=\hat{W}(f)+\frac{1}{NM}\sum_{i=1}^{N}\sum_{k=1}^{M}\hat{R}_{ik}(f),
\end{equation}

\noindent where $\hat{S}_{ik}(f)= \ln S_{ik}(f)$ and $S_{ik}(f)$ is the Fourier transform of $s_{ik}(t)$.

The estimated wavelet $\hat{W}_{e}(f)$ converges to the true wavelet $\hat{W}(f)$ if the reflectivity series is white, and the propagating wavelet is stationary. We refer to \citeasnoun{vanderbaan2010} for wavelet terminology.  Averaging zero mean random reflectivities over the ensemble of windows makes the real-valued part of the last term converge to a constant, and the imaginary part to zero. Hence the resultant process converges to the seismic wavelet. The complex log-spectrum of the averaged reflectivity is:
%\vspace{2mm}
\begin{equation}\label{eq:complxRefl}
\hat{R}(f) = \ln R(f)  = \ln |R(f)| + j \phi_R(f),
\end{equation}
%\vspace{4mm}
where the term $\ln|R(f)|$ converges to $\ln \sigma_r$ which approaches zero as reflectivity series have assumed unitary variance $\sigma_r$. The reflectivity phase $\phi_R $ tends to zero after deramping and phase unwrapping if we assume that (1) $\phi_R$ is uniformly distributed between $-\pi$ and $\pi$ and (2) the reflectivity series is dominated by a few large reflectors.

The recommended window length should be from three to five times the wavelet length, in accordance with \citeasnoun{Buttkus1975}. 

Our approach resembles the Welch transform in the way the seismic trace is divided into $M$ overlapping segments \cite{Welch1967}. We use overlapping windows to increase the number of traces thereby improving the estimation variance \cite{Angeleri1983}. The reduction in the estimation variance is of the order of MO compared to the use of the entire trace in one go (Appendix A). For a single seismic trace of length $T$, a window length $L$ and a fraction of overlap between windows $O$, the number of segments per trace will be:

\begin{equation}\label{eq:nosegments}
M=\frac{T-L O}{L-L O}.
\end{equation}

In Eq. (\ref{eq6:logspectral}) we use the complex logarithm. This means that phase information is included and we have to deal with the phase unwrapping problem \cite{Herrera2011}. Expanding Eq. (\ref{eq6:logspectral}) in terms of amplitude and phase:
\begin{equation}\label{eq:prod}
  \hat{W}_{e}(f) = \frac{1}{NM}\sum_{i=1}^{N}\sum_{j=1}^{M} \ln|S_{ik}(f)|+\frac{j}{NM}\sum_{i=1}^{N}\sum_{k=1}^{M} \arg\{S_{ik}(f)\},
\end{equation}

 \noindent where $j = \sqrt{-1}$ and the function $\arg$ refers to the continuous unwrapped phase $arg\{S(f)\}= \phi_S(f)+ 2\pi n$ with $n$ some integer. We need a continuous function to guarantee the uniqueness of the solution. That is why at every short trace we estimate the unwrapped and deramped phase. Removing the linear trend in the phase suppresses the wavelet timing prior to the averaging process, thereby improving the wavelet estimate.

 We tested the performance of various phase unwrapping algorithms \cite{Herrera2011} where we concluded that polynomial factorization \cite{Sitton2005} is exact but computational costly. On the other hand the $\omega$-plane method \cite{Kaplan2007} and the simple correction of $2 \pi$ jumps in phase \cite{Opp2010} have similar performance. Thus to reduce the computational load we remove all jumps of $2 \pi$ in the wrapped phase.

The estimated wavelet in the time-domain is finally given by:

\begin{equation}\label{eq:est_wavelet}
w_e(t)= FT^{-1}\{ \exp [\hat{W}_{e}(f)]\}.
\end{equation}

\section{Results}

To evaluate the performance of Short-Time Homomorphic Wavelet Estimation (STHWE), we compare its results with the log-spectral averaging (LSA) \cite{Otis1977,Tria2007} and with the true wavelet in the synthetic example and with the first arrival in the field dataset. We also include a kurtosis-based method to estimate a constant-phase wavelet \cite{VanderBaan2008}, as it gives an independent comparison based on an alternative approach using different statistical assumptions.

\subsection{Synthetic dataset}

\begin{figure}[ht]
 \begin{center}
  \includegraphics[scale=.7]{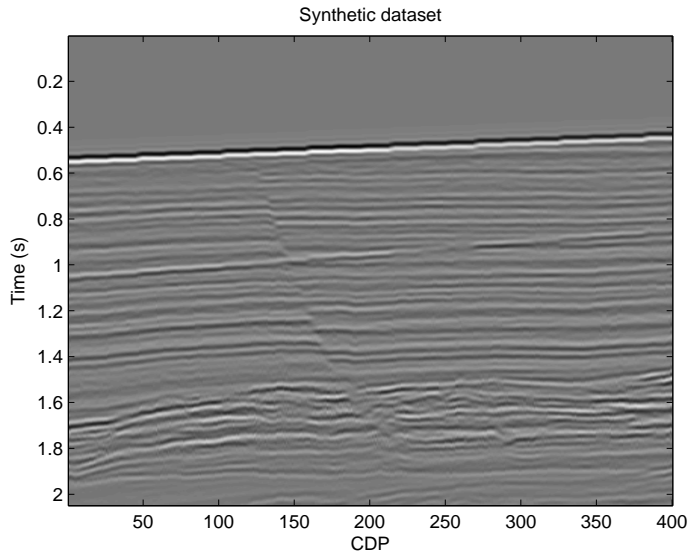}
 \end{center}
   \caption{Seismic dataset under study. Both statistical methods extract a global wavelet for the entire dataset.}
   \label{figR:dataset}
\end{figure}

The dataset consists of a synthetic section of 400 traces of 560 samples with 4 ms time step, see Figure \ref{figR:dataset}. These data have been produced with a mixed-phase and narrow-bandwidth Ricker wavelet. The dominant frequency range is between 9 and 37 Hz with a peak frequency of 23 Hz (i.e., 2 octaves and a ratio bandwidth to peak frequency of 1.18). Kurtosis-based methods require that the bandwidth exceeds the peak frequency \cite{Longbottom1988,White1988}.
The true wavelet has a complex frequency dependent phase with a mean value of 58 degrees. Inspection of the waveform of the strong first arrival shows that the wavelet is not zero phase as it is asymmetric. It has a large positive (black) and large negative (white) lobes.

Both statistical methods, KPE and STHWE are based on the Central Limit Theorem, thus averaging of traces is needed. The entire section is used to extract a single wavelet. The wavelet length was fixed to $wl = 220$ ms for both methods.

\begin{figure}[ht]
 \begin{center}
  \includegraphics[scale=.44]{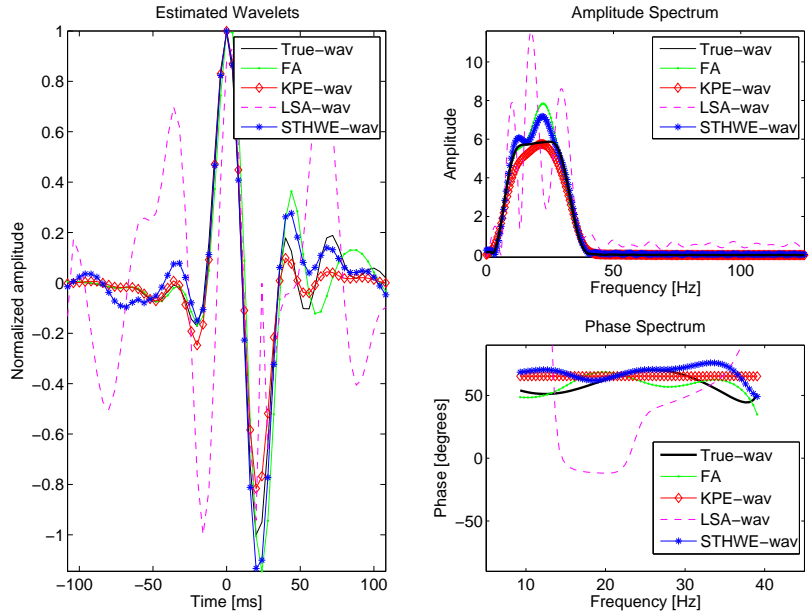}
 \end{center}
 \caption{Estimated wavelets. In the left in black is the true wavelet (True-wav), first arrival (FA) is plotted in green, the KPE method (red) and the short-time homomorphic method (blue). In the upper right panel the four amplitude spectrums are plotted and in right bottom their corresponding phase spectra. Average unwrapped phases are $\phi_{FA} = 58.42$ degrees, $\phi_{KPE} = 65.30$ degrees, $\phi_{LSA} = 49.06$ degrees  and $\phi_{STHWE} = 68.14$ degrees.}
\label{fig:waveletsA}
\end{figure}
The LSA method averages the entire trace while the short-time method uses small overlapping windows. In the STHWE method the critical parameter is the window length which guarantees the reduction of the estimation variance in the wavelet phase spectrum. We use an analysis window of 660ms (i.e., three times the wavelet length), tapered with a Hamming window, and set a 50 \% overlap between windows. Two cycles of averaging are implicit in our implementation, first one wavelet is estimated for each trace and the final estimated wavelet is the result of their average. For each trace we have $M = 19$ segments, obtained by the Welsh relation in Eq. \ref{eq:nosegments} and the total set is $NM = 7600$ segments. 
%The standard deviation of the reflectivity phase is $4.3$ degrees, which is the value affecting the wavelet phase estimation.

The true wavelet (black trace) is shown in Figure \ref{fig:waveletsA} along with the first arrival (FA in green), the KPE-wavelet(red), the LSA-wavelet (magenta) and the STHWE-wavelet (blue). The LSA estimated wavelet is unstable compared to the other three wavelets which all have similar time-domain waveforms (left panel) and spectral content.
LSA does not produce a very accurate result since it involves many more degrees of freedom (i.e., individual frequencies) compared with the STHWE method, combined with less averaging due to the longer windows. Finally we present the unwrapped and deramped phases for all the estimated wavelets. Their average phases are respectively $\phi_{KPE} = 65.30$, $\phi_{STHWE} = 68.14$, $\phi_{FA} = 58.42$ and $\phi_{LSA} = 49.06$, which are all close to 58 degrees, the average phase of the true wavelet.

\subsection{Noise Stability Test}
The noiseless synthetic example represent the optimum conditions for the proposed algorithm. In this section we explore the robustness of our approach in the presence of noise. We use the same dataset shown in Figure \ref{figR:dataset}, but contaminated with white Gaussian noise in a wide range of signal to noise ratios. Both methods, the kurtosis based estimation and the homomorphic wavelet estimation are tested using 50 Monte Carlo simulations per signal-to-noise ratio. All signal-to-noise ratios are defined as the standard deviation of the signal over that of the noise. 

For fairness of comparison, we also include a frequency dependent wavelet estimation method, called robust blind deconvolution (RBD) of \citeasnoun{VanderBaanPham2008}. This method is based on a modified mutual information criterion and operates on the amplitude and phase spectrum of the wavelet separately. The mutual information rate is a general-purpose criterion to measure
whiteness using statistics of all orders. It's principal parameter is the wavelet length which is set to 220 ms. The three wavelet estimates are correlated with the true wavelet, and the average of the correlation coefficients and their standard deviations are calculated. In each Monte Carlo simulation, different noise realizations are added to all traces. Here we follow the experimental setup by \citeasnoun{VanderBaanPham2008}.

Correlation coefficients are bounded by unity and therefore not normally distributed, which biases the estimated standard deviations. We use the Fisher's transformation to translate the correlation values to an almost normal distributed space. Then we compute the average and the upper and lower standard deviations. Finally, by inverse transformation we recover these values. See \citeasnoun{vandecar1990} for further details.

\begin{figure}[ht]
 \begin{center}
  \includegraphics[scale=.5]{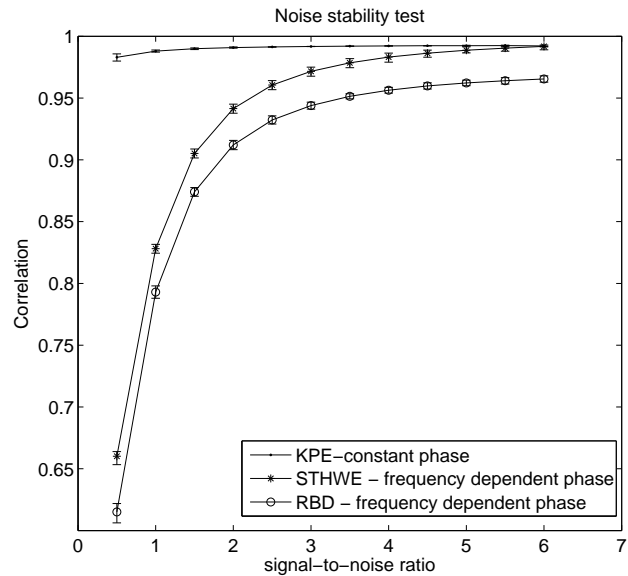}
 \end{center}
 \caption{Noise stability test. Signal-to-noise ratio versus correlation coefficient. The wavelet estimated by the constant-phase KPE and the frequency-dependent phase methods STHWE and RBD are compared to the true wavelet. The homomorphic method outperforms the other frequency-dependent technique; yet KPE is most appropriate here as this true wavelet can be accurately described by a constant-phase approximation.}
\label{fig:noisetest}
\end{figure}

Figure \ref{fig:noisetest} shows the correlation coefficients between the estimated wavelets and the true wavelet. All three wavelet estimation techniques perform well for high signal-to-noise ratios. The KPE method performs best since the compact nature of the true wavelet favors the constant phase estimation. KPE achieves, in this case, better reconstructions as seen by the higher correlation coefficients, but \citeasnoun{VanderBaanPham2008} show that complex wavelets such as dispersive ones cannot be described by a constant phase approximation. Therefore, the two frequency-dependent methods are most useful for more
band-limited data e.g., old legacy data and situations where long wavelets with complex phases are anticipated.

The homomorphic method outperforms the RBD method, giving satisfactory result for signal-to-noise ratios over 1.5. The three methods have low standard deviations in their estimates at all noise levels. 

\subsection{Real dataset}
\begin{figure}[ht]
 \begin{center}
  \includegraphics[scale=.68]{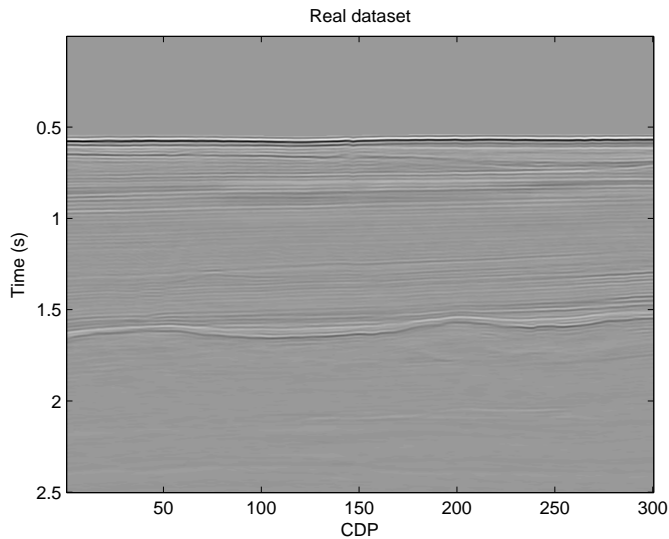}
 \end{center}
 \caption{Real stacked seismic section. The first arrival corresponds to the ocean-bottom reflection.}
\label{fig:realdataset}
\end{figure}

Finally, we show the performance of the new wavelet estimation method on a real seismic dataset. A stacked section of marine data is shown in Figure \ref{fig:realdataset}. The original section consists of 300 traces and 1074 samples, with a sampling frequency of 250 Hz. The first arrival occurs around 0.55 s, the ocean bottom. There is a strong reflector at approximately 1.6~s, below which few arrivals are observed. In real data the wavelet length can often be selected from observation of the length of the first arrival, in this case $wl = 140$ ms. The analysis window is set to be three times the wavelet length, using a 50 \% overlap and a Hamming filter for tapering.

\begin{figure}[ht]
 \begin{center}
  \includegraphics[scale=.44]{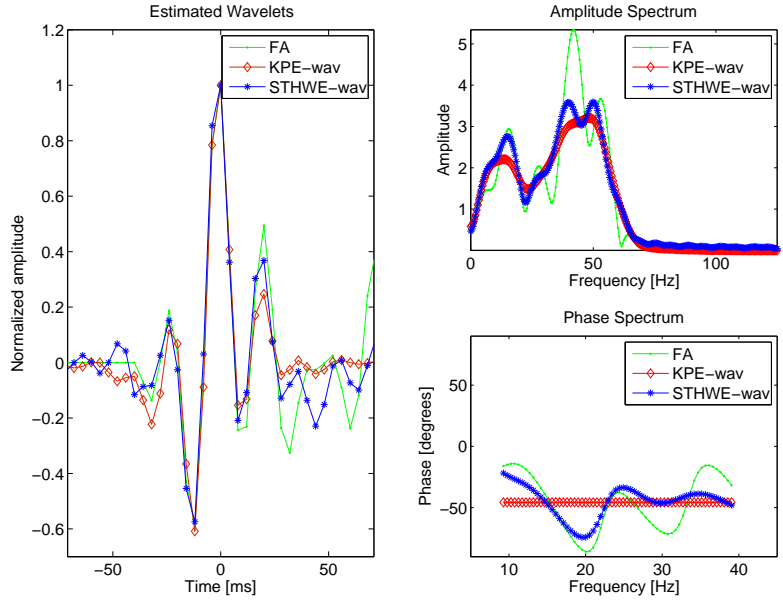}
 \end{center}
 \caption{Estimated wavelets for the stacked section displayed in Figure 4. The constant-phase rotation approach (KPE) in red, and the new short-time homomorphic method (blue) lead to similar wavelet estimates, with high similarity to the first arrival (FA) in green. The KPE amplitude spectrum is smoother than the homomorphic estimates due to its constant-phase nature. The KPE phase estimate is $\phi_{KPE} = -45$ degrees, whereas the average phase for the STHWE and first arrival are respectively $\phi_{STHWE} = -45.13$ and $\phi_{FA} = -45.97$ degrees.}
\label{fig:realwavelets}
\end{figure}

The estimated wavelets using the constant-phase rotation approach (KPE) and the homomorphic method (STHWE) are compared with the first arrival (FA) at 0.5 s. Figure \ref{fig:realwavelets} shows in the left plot the estimated wavelets  by the constant-phase rotation approach (red) and the new short-time homomorphic method (blue). Both estimates lead to similar wavelets and reproduce the FA waveform, as expected in this marine example with a strong ocean-bottom reflection.

The mean value of the frequency dependent phase for the STHWE method and the first arrival are respectively $\phi_{STHWE} = -45.13$ and $\phi_{FA} = -45.97$ degrees. The KPE method delivers a constant phase of $\phi_{KPE} = -45$ degrees. Thus, both statistical methods produce similar results comparable to the first arrival.

\section{Discussion}

The minimum phase constraint on the reflectivity function is a fundamental assumption in the technique of \citeasnoun{Ulrych1971}. It is also a necessary condition for full trace log-spectral averaging proposed by \citeasnoun{Otis1977}. In the proposed short-time averaging method it is no longer required as many reflections are taken into account. Phase unwrapping and deramping tends to place the largest reflection at low quefrency values, thus allowing the repeated averaging to estimate the underlying propagating wavelet.

The sparsity condition, in the conventional homomorphic deconvolution method, aims to separate the wavelet from the reflectivity in the cepstral domain. This means that the interarrival time should be sparse enough to isolate the wavelet from the reflectivity. As the time separation between arrivals is unpredictable, induced whitening due to applying a logarithm to the wavelet passband may not be sufficient to separate individual arrivals. In our case a whiteness assumption of the reflectivity leads to random rhamonics in the log-spectrum, which after averaging converge to the smooth wavelet log-spectrum, i.e., the averaging process acts as a lifter.

The stationary condition imposed on the wavelet, is a limitation in time varying wavelet estimation. In this case, we suggest to assume that the wavelet is piecewise stationary and divide the seismic section into two or more separate time windows to estimate different wavelets.

\section{Conclusions}

Homomorphic wavelet estimation was first introduced over 40 years ago and has been revisited often with its promise of nonminimum-phase wavelet estimation. The original method of cepstral liftering assumes the wavelet has a smooth spectrum and that the reflectivity series is minimum phase and sufficiently sparse. However, in most of the cases, the latter assumption is rarely honored.

Log-spectral averaging mitigates the need for the sparsity constraint, but requires a large number of independent reflectivity series while maintaining the minimum phase constraint. The method of log-spectral averaging using a short-term Fourier transform increases the number of traces, thus reducing estimation variances. Furthermore, no assumptions regarding the phase of the wavelet or the reflectivity are required.  A comparison using a synthetic example shows similar results, with regards to constant-phase wavelet estimation based on kurtosis maximization. The short-time homomorphic method and the kurtosis-based method produce similar wavelets but the short-time homomorphic technique allows for a frequency-dependent phase estimation, whereas the kurtosis-based method assumes a constant phase.

The noise stability test, using Monte Carlo simulations on a synthetic gather, demonstrates the feasibility of statistical wavelet estimation from noisy seismic traces. 

Both statistical methods lead to reliable wavelet estimates that could be employed as quality control tool for deterministic methods. When different well logs in the same seismic section produce different extracted wavelets, the statistical methods could interpolate the wavelet phase between the non-matching wells. Seismic inversion in absence of well logs is one the main applications for the statistical methods. The method also shows promise for estimating wavelets from time-lapse datasets where the acquisition parameters may have changed between monitoring and baseline surveys or significant time-lapse changes are observed around key horizons.

\ack The authors thank Chevron for permission to use the synthetic data and BP for providing the real dataset and the Sponsors of the BLISS project for their financial support. We are grateful to the two anonymous reviewers for their valuable comments and suggestions to improve the original manuscript. R. Herrera thanks Mauricio Sacchi for his useful comments related to the Homomorphic Deconvolution Method.

\appendix
\section{Phase variance in the short-time homomorphic analysis}

The quality of the estimated wavelet log-spectrum is highly dependent on the number of averaging windows. Better results are obtained with a larger number of averaged segments for a stationary wavelet. The objective is to minimize the effect of the reflectivity in the second right term of Eq. \ref{eq:prod} for the time-invariant case.

\citeasnoun{Angeleri1983} evaluates the number of windows needed for a successful averaging process.
Denoting the variance of the phase spectrum of the averaged reflectivity by $\sigma_R^2$ the original expression is:
\begin{equation}\label{eq:varianceAng}
 \sigma_R^2=\frac{\pi B L }{12 N},
 \end{equation}
 
\noindent where $B$ is the frequency bandwidth, $L$ is the trace length and $N$ is the total number of traces.

Assuming statistical independence between non overlapping segments it is reasonable to assume we can approximate the averaged reflectivity as:

 \begin{equation}\label{eq:variance}
 \sigma_R^2=\frac{\pi B L }{12 NMO},
 \end{equation}

\noindent with $O$ the fraction of overlap in each segment and $M$ the total number of segments. Using short-time windows reduces therefore the estimation variance \cite{Ulrych1995}. Note that the dimensionless value of $\sigma_R^2$ refers to variance of the phase in radians.

\section*{References}

\bibliographystyle{jphysicsB}
\bibliography{blisshomomorphic}

\end{document}